# The Weather and its Role in Captain Robert F. Scott and his Companions' Deaths

Krzysztof Sienicki

*Chair of Theoretical Physics of Naturally Intelligent Systems*
*ul. Topolowa 19, 05-807 Podkowa Leśna, Poland, EU*
krissienicki@yahoo.com
24 October 2010

## Abstract

A long debate has ensued about the relationship of weather conditions and Antarctic exploration. In no place on Earth is exploration, human existence, and scientific research so weather dependent. By using an artificial neural network simulation, historical (Heroic Age) and modern weather data from manned and automated stations, placed at different locations of the Ross Ice Shelf, and the Ross Island, I have examined minimum near surface air temperatures. All modern meteorological data, as well as historical data of various sledging parties, Cherry – Garrard data, high correlations between temperatures at different locations, and artificial neural network retrodiction of modern and historical temperature data, point out the oddity of Captain Scott's temperature recordings from February 27 – March 19, 1912. I was able to show that in this period the actual minimum near surface air temperature was on the average about 13°F (7°C) above that reported by Captain Scott and his party. On the basis of the mentioned evidence I concluded that the real minimum near surface air temperature data was altered by Lt. Bowers and Captain Scott to inflate and dramatize the weather conditions.

## 1. Introduction

After months of anticipation and high hopes the world learned that Captain Scott and his four companions perished on their way back from the South Pole. Something that was unimaginable happened [4,14]. An official cable communiqué from Christchurch [5], New Zealand on February 10, 1913, tells us that; 'Scott, Wilson and Bowers, died from exposure and want during a blizzard about March 29'. One can fairly assume that this announcement was based on the weather data from Captain Scott's diaries and the descriptive account in his famed and extraordinary *Message to the Public* '… no one in the world would have expected the temperatures […] which we encountered at this time of year'[14]. The British Antarctic Expedition 1910-1913 and Captain Scott's role has been analyzed from many points of view [4]. Eleven years after the tragic events in the Antarctic, expedition meteorologist, Dr. George G. Simpson explained during his Halley Lecture [7] that; 'Whatever other causes there may have been, there can be no doubt that the weather played a predominant part in the disaster and […] was the intermediate cause of the final catastrophe'. Another member of the expedition, Charles S. Wright also analyzed temperature data. He observed that minimum temperatures reported by Captain Scott's party during their final weeks were normal [24]. However, Wright stressed that the data available to him in 1974 were far from satisfactory. He stated that, "It will probably not be long before more and better information becomes available since unmanned meteorological stations have become available." The first automated weather station at the proximity of Captain Scott's route at the Ross Ice Shelf started to transmit data in 1985. Since that time a continuous record has been available. In the meantime neural network methods of time series data analysis had been expertly developed. It appears that the present record of 25 years of data combined with advanced methods of analysis, may shed a new light on present and past meteorological events.

## 2. Data

In this paper I will use two sets of daily *minimum* near surface air temperature data measured at various geographical locations at Ross Island and at the Ross Ice Shelf (Fig.1). These sets of data were collected during the years 1911-1912 and 1985-2009. In the first case, temperature data were measured by the members of the British Antarctic Expedition 1910-1913 (*Terra Nova*) under the command of Captain Robert F. Scott. In the second case, respective minimum near surface temperature data were measured by modern automated weather stations.

### 2.1 British Antarctic Expedition Historical Temperature Data (1911-1912)

One of scientific aspects of Antarctic exploration was the collection of meteorological data to be used in validation and further development of the understanding of global air circulation. All pertinent aspects of meteorological measurements, data and discussion were collected in a three volume treatise by George Simpson [18]. He was *Terra Nova* expedition's chief meteorologist. Historical temperature data constitute a sub-set of meteorological records. These records were collected at permanent weather stations and at the routes travelled by Captain Scott and various auxiliary sledge parties. Approximate locations of land weather stations (historical and modern) are clearly depicted on Fig.1. During the *Terra Nova* expedition the land based meteorological measurements were taken at Cape Evans of Ross Island. The measurements were taken every hour by Simpson and his Canadian assistant Charles Wright. The temperatures were taken in the screen mounted behind the expedition hut about five feet above ground on *Windvane Hill* [18]. Four thermometers were placed in the screen: a mercury dry bulb thermometer, a mercury maximum thermometer, a spirit minimum thermometer and a bimetallic thermograph. The measurements were taken in air-free conditions. Before and after the expedition, Simpson ensured adequate testing procedures [18].

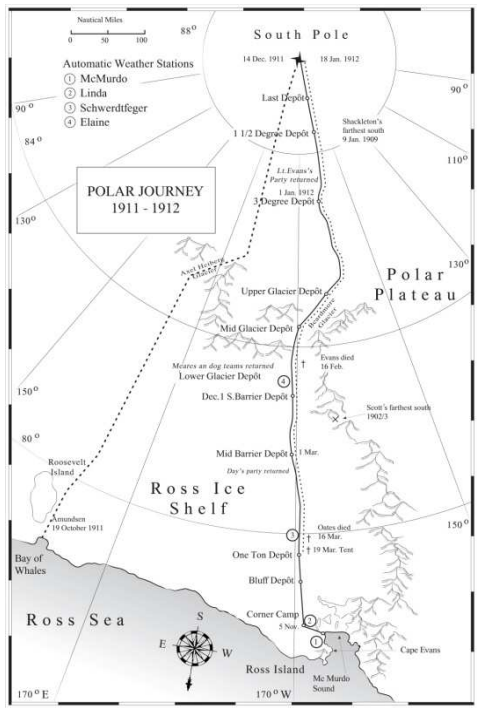

Figure 1. Approximate drawing of the Antarctic route travelled by Captain Scott (solid and doted lines), and the auxiliary and relief parties in 1911 and 1912, which travelled essentially the same routes as Captain Scott's party. Positions of automatic weather stations are also shown. See figure legend for additional information.

The sledging parties did not take hourly measurements for obvious reasons. Usually three measurements were taken: in the morning, at lunch time and in the evening. Some sledging parties, for example the Main Polar Party, were carrying so-called minimum temperature thermometers, in addition to regular thermometers. Captain Scott's party used high quality thermometers, calibrated at Kew Observatory, London. Sling and dry-bulb thermometers were used with precision, and measured at about a $\pm 0.5°F^1$ uncertainty [18]. A specially constructed sling thermometer with a wooden handle was broken by Lt. Bowers on March 10, 1912. From that day on, only Captain Scott's personal spirit thermometer data were available.

## 2.2 Modern Temperature Data (1985-2009)

Each automated weather station measures wind speed, wind direction, temperature, and atmospheric pressure. The wind speed, wind direction, and temperature gauges (sensors) are mounted at the top of the tower, at a nominal height of 3.9 m. Station atmospheric pressure is measured at a nominal height of 1.5 m. The heights of the gauges (sensors) may change due to snow accumulation at the site. Measurements from the sensors are made every 10 minutes and are transmitted *via* the ARGOS data collection system and processed at the University of Wisconsin. A semi-automated quality control process is applied to 10-minute data. Untreated data are also available. Hourly observations are created using the closest valid observation within 10 minutes of the hour from the quality control processing.

As a reference I used modern meteorological data provided by the British Antarctic Survey and the Antarctic Meteorological Research Center, the University of Wisconsin for the following weather stations (in decimal degrees):

McMurdo (-77.85°, 166.667°)       - Altitude: 24 meters,
Linda (-78.480°, 168.375°)        - Altitude: 50 meters,
Schwerdtfeger (-79.904°, 169.97°) - Altitude: 60 meters,
Elaine (-83.134°, 174.169°)       - Altitude: 60 meters.

In the analysis I have used the full record of temperature data for the McMurdo and Schwerdtfeger automated weather stations for the years 1985-2009. I used the full record of temperature data from the Linda and Elaine stations for the years 1993-2009, including March 2009 data [6]. Due to some temporary failures of the automated weather stations or also due to satellite data transmission problems, the available data (10-minuta data) were treated with spline functions (with added Brownian noise as the null hypothesis) to fill minor omissions and small gaps. I could have used other methods [13] for the same purpose, however, the spline technique works just as well and is faster. This method was not used for records of minimum near surface air temperatures.

---

[1] In this paper, in order to be consistent with the historical account, I will use Fahrenheit instead SI Celsius temperature units. The relationship between Celsius and Fahrenheit temperature scales is $[°C] = \frac{5}{9}([°F] - 32)$.

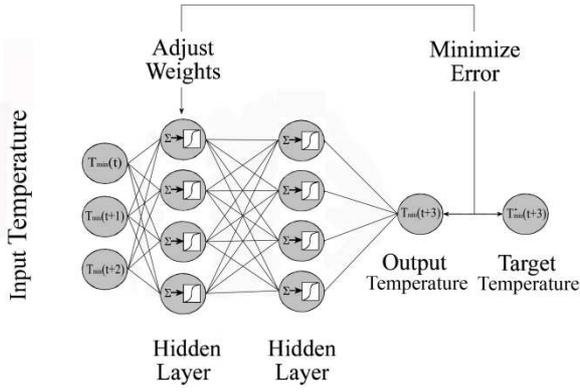

Figure 2. Architecture of back-propagation artificial neural network used in this work.

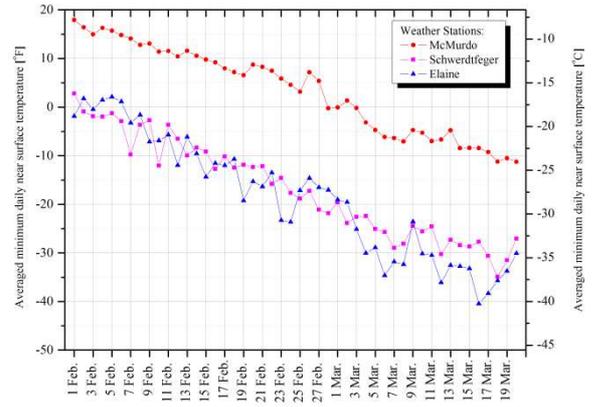

Figure 3. The minimum daily near surface temperature averaged (the arithmetic mean) from the 1985-2009 McMurdo, Schwerdtfeger and 1993-2009 Elaine weather stations

## 3. Methods

To perform retrodiction of the daily minimum temperature at the geographical locations of the weather stations, I have selected a back-propagation neural network. It is a network which is able to train the output (Linda, Schwerdtfeger, Elaine minimum temperature data) units to learn to classify patterns of inputs (McMurdo minimum temperature data) [12]. There is no theoretical prescription for the number of hidden layers. I have found that a fully connected back-propagation neural network as depicted on Fig. 2 performed very well for the temperature data studied in the present paper. The neural network used in this paper was a modified version of the neural network used by me in previous studies of solutions of the quantum mechanical Schrödinger equation [1]. While experimenting further with different configurations of the network I have found that a network with 3 input neurons and 4 neurons in each first and second hidden layer and 1 output neuron, gave the best retrodiction results. The neural network was trained on sets of daily minimum temperatures in the following way. I presented to the input neurons, a sequences of equally spaced samples of McMurdo daily minimum temperatures $T_{min}(t)$, $T_{min}(t+1)$, $T_{min}(t+2)$, where t stands for a particular Julian day considered in this work period of time. The output neuron was assumed to be the desired retrodiction daily minimum temperature $T^*_{min}(t+3)$ from the respective Linda, Schwerdtfeger and Elaine data. After the network reached its maximum performance for the given set of data, I shifted $t \to t+1$ and repeated the learning procedure. Thus, in such a way I have obtained three fully trained neural networks for the retrodiction of minimum daily temperatures at geographical coordinates of the mentioned weather stations.

The mean *absolute* retrodiction error $\langle\varepsilon\rangle$ was calculated from

$$\langle\varepsilon\rangle = \pm\frac{1}{N}\sum_{j=1}^{N}\frac{1}{n}\sum_{i=1}^{n}|T_{AWS}^{(i,j)} - T_{ANN}^{(i,j)}|$$

where N is the number of years, n is the number of days. The minimum daily temperatures at automated weather station (AWS) and retrodicted by artificial neural network (ANN) are denoted respectively. The error $\langle\varepsilon\rangle$ is not the standard deviation of the sample of retrodicted minimum temperatures. It is the mean *absolute* retrodiction error.

Automated weather stations are localized in the proximity of Captain Scott's route and expedition depots: Corner Camp (Linda), One Ton Depôt (Schwerdtfeger), and South Barrier Depôt (Elaine) as depicted in Fig.1.

After selecting architecture for the best performing neural network I have examined its retrodiction performance by sequential deselecting of yearly minimum temperature data from the training data series. This was done for the McMurdo, Schwerdtfeger stations and performed retrodiction of deselected data was made. Artificial neural network performed extremely well in retrodiction of the minimum temperatures for 25 consecutive years of modern data. The performance of the artificial neural network is described in the following section. I have calculated the mean absolute retrodiction error of the sample of retrodicted minimum temperatures from February 27 to March 19 for each year. The result being $\langle\varepsilon\rangle = \pm 7.1°F$. This indicates a fairly acceptable (see Results and Discussion) precision of the retrodiction power of my artificial neural network.

## 4. Results and Discussion

In this paper, for the sake of clarity, I have presented a detailed analysis of the daily minimum near surface air temperatures for historical [18] and modern data [6]. The approximate positions of the automated weather stations and travelled route by different parties are depicted in Fig. 1. For analysis I have selected four weather stations which are in proximity to the routes travelled by Captain Scott and his auxiliary parties, and relief parties. McMurdo station is situated at the Hut Point Peninsula of Ross Island about 26 km from Captain Scott's Hut at Cape Evans. This is where Dr. Simpson and his assistant Charles Wright took weather data measurements all year round from 1911-1912. In the analysis I have used the full record of temperature data for the McMurdo and Schwerdtfeger stations for the years 1985-2009. I used the full record of temperature data from the Linda and Elaine stations for the years 1993-2009, including March 2009 data [6].

Figure 3 depicts the averaged daily minimum temperature change from February 1 to March 19, at the McMurdo, Schwerdtfeger and Elaine weather stations. Even though these two latter stations are approximately 229 and 588 km apart from McMurdo, one can easily notice that some unspecified relationship of minimum temperature changes between these stations is present. A gradient in the minimum temperature at one station is followed by a similar change at another station and/or *vice versa*. Thus, the components of the gradient of minimum daily temperature, transform covariantly under changes of coordinates i.e., the geographical coordinates at the Ross Ice Shelf. I do not imply that there is a mathematical rigorousness of the mentioned covariant transformation and/or linear relationship.

Although averaged minimum temperatures presented in Fig. 3 can be exceptionally well approximated by linear regression analysis, its very nature is nonlinear. One has to notice that I have presented only temperatures for the first quarter of the year. The actual yearly averaged minimum temperature has a U shaped letter with a distinctive coreless winter, April through September and a short-lived crest temperature between the beginning and end of summer, December through January. Fig. 3 also shows that although the Elaine station is further South than the Schwerdtfeger station, the daily minimum temperature at the Elaine is frequently close to or above that of the Schwerdtfeger station. I attribute this phenomenon to an adiabatic effect of air warming by katabatic winds flowing downwards from the Beardmore Glacier. With this in mind one can further confirm a mirrored similarity of the daily minimum temperatures changes between McMurdo, Schwerdtfeger, Elaine and Linda (which is not shown for clarity) weather stations. I attribute this "*mirrored similarity*" to the essentially flat surface of the Shelf and to the prevailing south and south-east by south winds that are directed along the Transantarctic Mountain pathway. Therefore, a mirrored similarity of minimum temperatures along the route of Captain Scott and the auxiliary parties is self evident.

I could have advance my study by using the above mirrored similarity. It seemed essential in the analysis, however, to account for fluctuations and nonlinear trends of minimum temperature changes as a continuous-state, discrete-time stochastic process. For this purpose I have selected an artificial neural network for a time series prediction and retrodiction of minimum temperature data. After selecting architecture for the best performing neural network I have examined its retrodiction performance by sequential deselecting yearly minimum temperature data from training data series. This was done for the McMurdo -Schwerdtfeger stations and performed retrodiction of deselected data was made. From Simpson's experimental studies and report we know that the uncertainty of thermometers used by Captain Scott's expedition was about ±0.5°F [18]. It is roughly the size of points indicating temperatures changes in the figures in this paper. From a possible 25 year set of training temperature data for the McMurdo-Schwerdtfeger stations I have selected a 24 year set of training data and performed retrodiction for the not selected for training 25th year. Thus for each deselected year I calculated the respective *absolute* retrodiction error for retrodicted minimum near surface temperatures, for the days from February 27 until March 19. In this way I have obtained *absolute* retrodiction error for each deselected year; all together 25 *absolute* retrodiction errors. The mean *absolute* retrodiction error is $\langle \varepsilon \rangle = \pm 7.1$°F.

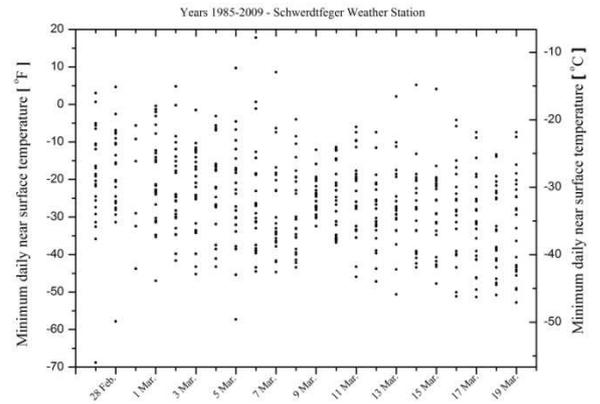

Figure 4. The minimum daily near surface air temperatures recorded at Schwerdtfeger weather station during the 1985 - 2009 period.

In order to illustrate variability of minimum near surface temperatures and the capacity of the artificial neural network which is used in this work, to learn and retrodict widely fluctuating temperatures, I have presented the fluctuations of minimum near surface temperatures at the Schwerdtfeger station for all years available on Figure 4. One can readily notice that minimum near surface temperatures, fluctuate in a wide range <+17.8, -68.8>°F of possible values. Taking into account the fact that the near surface air minimum temperatures at the Schwerdtfeger weather station fluctuate between $\langle +17.8, -68.8 \rangle$°F, this indicates a fairly acceptable precision: about $\langle \varepsilon \rangle = \pm 7.1$°F of the retrodiction power of my neural network.

Historical near surface minimum (or the *lowest*) daily temperatures were reported by: the Main Polar Party (Nov. 14-22, 1911), the Motor Party (Nov. 8-13 & Dec.1-9, 1911), the Day's Depot Party (Jan. 7-11, 1912), the First Return Party (Jan.17,18, 1912), the Second Return Party (Feb.4-9, 1912)

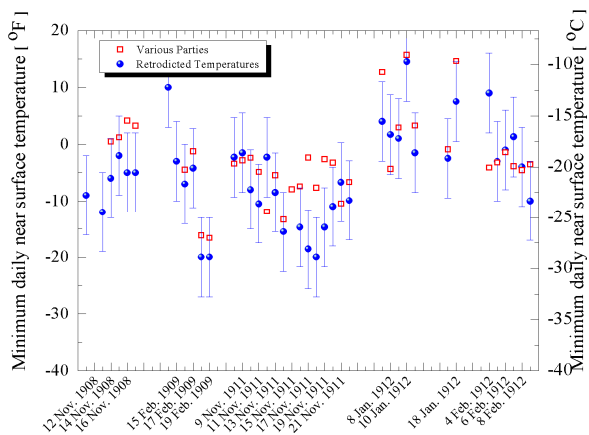

Figure 5. Historical minimum (or the lowest) and retrodicted daily temperatures at envious of One Ton Depôt.

[18] and the Shackleton Party (Nov.11-17, 1908 & Feb.11-19, 1909) [10]. Minimum daily temperatures were retrodicted from historical minimum temperature data recorded at Cape Evans and Cape Royds respectively. It is evident from these simulations, that my neural network is capable of precise retrodiction of all minimum temperatures reported by various parties at different times of the year. None of the measured and retrodicted minimum temperatures are biased. As one would expect, they randomly fluctuate around each another. From the above, I have established and verified essential pieces of evidence with which I can retrodict with accuracy, respective temperatures at environs of One Ton Depôt based on minimum temperature measurements at: McMurdo (1985-2009), Cape Evans (1911-1912) or Cape Royds (1908-1909). I turn now to the analysis of minimum temperature data reported by Captain Scott's Main Polar Party at the end of their arduous journey in February and March 1912.

The measured and reported [14,18] minimum temperatures by Captain Scott's party together with retrodicted values in the vicinity of the Schwerdtfeger weather station (One Ton Depôt, the final miles and the last of Captain Scott's camps) form the historical data of minimum temperatures measured at Cape Evans. They are clearly depicted in Fig.6. It is not difficult to note that until February 27, 1912 the retrodicted and the temperatures reported by Captain Scott confirm anticipated short lived random fluctuations. Compared to the temperature data presented in Fig. 3 for the Schwerdtfeger and Elaine stations, Captain Scott's reported minimum temperatures were slightly above retrodicted temperatures for the geographical position of One Ton Depôt. This is to be expected as his party was South from that depôt between February 15-23. It should also be noted that until the latter part of February the formerly observed (Fig. 3) mirrored similarity, temperature reflection was well established.

However, from February 27 on, for twenty consecutive days (with an exception of March 4, see discussion below) reported and retrodicted minimum daily temperatures rapidly and significantly diverge. Retrodicted minimum near surface temperature, is on the average -13ºF above that reported by Captain Scott and his party.

The discrepancy is significant in value and length. Such a discrepancy can be attributed to:
(i) inaccuracy of my retrodiction method,
(ii) incorrect temperature data readings due to Captain Scott's party thermometers malfunction,
(iii) long term temperature change in the Antarctic and/or
(iv) the fact that temperature data documented by Lt. Bowers and Captain Scott were distorted to exaggerate real weather conditions.

In the above analysis and discussion I have clearly confirmed that I was able to retrodict historical minimum temperature data reported by various parties. Narrowing down to the weather conditions at One Ton Depôt, it can be argued that because my data for the training of a neural network at the Schwerdtfeger station did not contain freak temperature events, I could not retrodict similar events in the past. However, this is not the case. The training set of temperature data contained all possible temperature recordings (see Fig. 4) and the observed minimum temperatures fluctuated in the wide range of $\langle +17.8, -68.8 \rangle$°F. Therefore, I would argue that all minimum temperatures, except Captain Scott's temperature data, measured at the environs of the final camp, show that the Main Polar Party reported temperature data that are in dispute with all the remaining temperature data. In the case of the Schwerdtfeger station, for retrodiction I have used twenty five years (1985-2009) of solid temperature data which contain a great number of minimum temperature fluctuations. (One more time see Fig. 4.) More importantly, only temperature data from Captain Scott and his party for late February and March diverge from the established trends and do not reflect the *mirrored similarity* of minimum temperatures measured at Cape Evans by Simpson and Wright. It is also pertinent to note that it is a remarkably long time to have twenty consecutive days of divergence in minimum temperatures. This alone excludes the commonly held notion of a rare event occurring in the system which constantly fluctuates around its mean. In my case it is the arithmetic mean of minimum daily near-surface air temperature.

I have already mentioned that in training and followed retrodiction calculations I have used temperature data collected at the McMurdo weather station. I have pointed out that this station is about 26 km from Captain Scott's Hut at Cape Evans where actual historical measurements were taken. Although there is no satisfactory historical or modern evidence [18] one could imply that there could be a possibility of temperature difference between these two sites. My neural network is ideally suited to answer any question as to how eventual temperature difference would affect overall prediction and retrodiction procedures related to discussed case. I have performed a numerical experiment by training a neural network with minimum temperature data different from original McMurdo data by ±3.6ºF (±2ºC). After that I have used his-torical data to retrodict minimum temperatures at One Ton Depôt. The results are depicted on Fig. 6. The upper dashed curve was obtained after the McMurdo data were perturbed by +3.6ºF and the lower dashed curve by -3.6ºF, respectively. This numerical experiment shows that even if there was slight temperature dependence due to intricate meteorological and/or physical features at Ross Island it was negligible and insignificant in formulating the conclusions of the present work. Simultaneously this numerical experiment further confirms that the artificial neural network used in this work is sensitive and capable of response to fine temperature changes in training and analysis data.

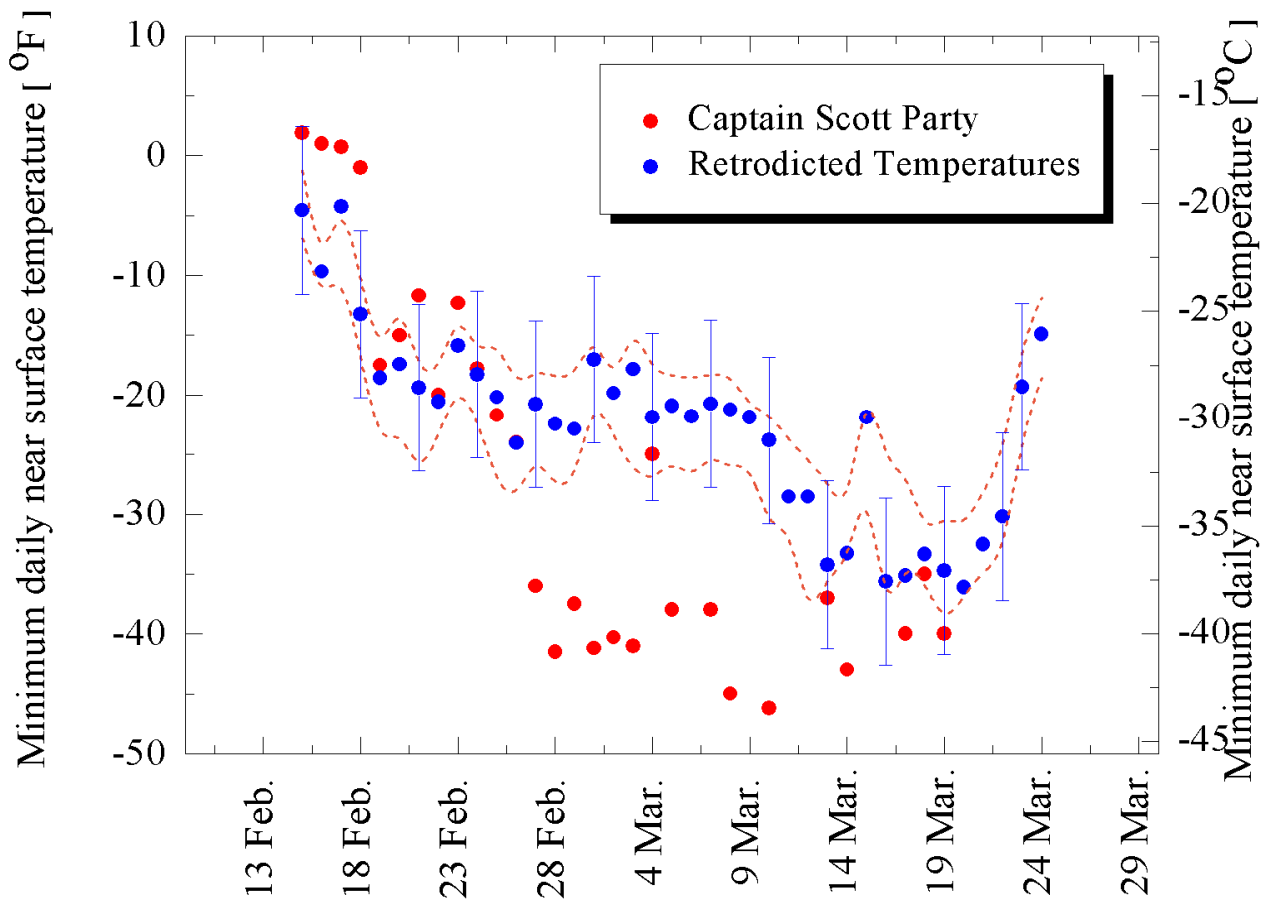

Figure 6. Historical near surface minimum daily temperatures (or the *lowest*) reported by Lt. Bowers and Captain Scott together with respective minimum temperatures retrodicted from historical minimum temperatures recorded at Cape Evans in 1912. Temperature readings after March 10 *are not minimum* near surface temperatures. See text for more discussion.

I have already mentioned that the uncertainty of thermometers used by Captain Scott's expedition was ±0.5ºF [18]. The only thermometer left after March 10, 1912 was Captain Scott personal spirit thermometer which was found by a search party in 1913. Charles Wright tested its calibration back in London. Test results proved this thermometer's accuracy within a tenth of a degree [19, pp. 289].

Although, scientific debate over the issue of global warming is still open, recent results further confirm a positive warming trend in the Antarctic [20]. The warming trend varies with the geographical position over the Antarctic, however for approximate coordinates of the Schwerdtfeger weather station the estimated warming trend is about 0.18ºF (0.1ºC)/decade. I believe that this warming trend does not contribute significantly to my overall analysis and conclusions.

By eliminating the alternative explanations (i), (ii) and (iii) mentioned above, only one conclusion is left (iv). That is, that the temperature data reported by Lt. Bowers and Captain Scott himself in late February and March 1912 were distorted by them to exaggerate and dramatize the weather conditions. My results clearly show that Captain Scott, Dr. Wilson and Lieut. Bower's deaths were a matter of choice rather than chance. The choice was made long before the actual end of food, fuel and long before the end of their physical strength to reach imaginary salvation at One Ton Depôt.

The March 4[th] minimum near surface temperature reported by Captain Scott is much higher than all remaining temperatures between February 27 until March 19. Coincidently, the March 4[th] temperature spike is very close to the retrodicted by me temperature for this particular day. On this particular day Bowers recorded: 5am/-24.3ºF, 1pm/-13.5ºF, 9pm/-25ºF and minimum temperature -24ºF ([18], vol. III, pp. 641). But a sudden increase in temperature started the day before: 5am/-41ºF, 1pm /-13.8ºF, 7.30pm/-15.5ºF. The rise in temperature by 27°F on March 3rd within 8 h without change of wind direction and its speed is an *extraordinary* meteorological event in Antarctic, especially from the perspective of Captain Scott's worn down party. And yet neither of the party including Captain Scott, comment on it on that day. Only on the next day (March 4th) Captain Scott casually adds in his journal "For the moment the temperature is -20°, an improvement which makes us much more comfortable, but a colder snap [*sic*] is bound to come again soon." The phraseology of this ruling and its logic is an interesting one, especially if one remembers Captain Scott's famed *Message to the Public* and the explanatory line "no one in the world would have expected the temperatures and surfaces which we have encountered at this time of the year."

Until automated weather stations were installed no meteorological data were available for the analysis of all-year round weather conditions on the Ross Ice Shelf. However, in spite of that, Heroic Age Antarctic explorers as well the scientists accompanying them, by assuming spatiotemporal continuity of meteorological events, could gain insight into the question. James Murray, a biologist of Ernest Shackleton *Nimrod* expedition (1907-09) used weather data from all previous Antarctic expeditions and produced interesting figures of yearly change in costal temperatures. This summary was known to Simpson who together with Captain Scott and Dr. Wilson, while planning the South Pole attempt, guessed with fair accuracy possible temperatures on the Ross Ice Shelf during expedition time[18]. Their analysis was summarized by Simpson in the first volume of *Meteorology* of the *Terra Nova* expedition. The estimated *difference* in the mean temperature at McMurdo Sound (Cape Evans) and One Ton Depôt was about -21.4°F for the month of March [18]. Knowing the mean temperature at Cape Evans for March, which was about +4.4°F, [18] Captain Scott should have expected the mean temperatures for the entire month of March to be about -17°F in the proximity of One Ton Depôt. Pending a steady seasonal gradient of temperatures (see Fig. 3) one would anticipate slightly higher temperatures at the end of February and early March. None of that happened. On the contrary. There was a never ending cold snap and cold outbreak. The rise of temperatures on March 3rd was "an improvement" as Captain Scott noticed but not a return to expected air temperatures. Captain Scott's description of meteorological events in the *Message to the Public* and actual events described in his journal at the beginning of March 1912 are somehow contradictory. In the *Message* Captain Scott informs the public that "no one in the world would have expected the temperatures" but in the journal entry on March 4th Captain Scott *predicts* that after a short warming on March 3rd and 4th "a colder snap is bound to come again soon." This discrepancy is puzzling. Captain Scott is talking without speaking.

On another occasion, almost exactly three months earlier, just at the entrance to the Beardmore Glacier, Captain Scott's South Pole party was stopped and held in the tents for four long days by blizzard or gale as he called it in the *Message to the Public*. Captain Scott asks, "What on earth does such weather mean at this time of year?" and wonders [14]:

> "Is there some widespread atmospheric disturbance which will be felt everywhere in this region as a bad season, or are we merely the victims of exceptional local conditions? If the latter, there is food for thought in picturing our small party struggling against adversity in one place whilst others go smilingly forward in the sunshine."[2]

Figure 3 provides the answer in that there are high correlations between minimum near surface temperatures along the western Ross Ice Shelf. There is a less pronounced correlation of wind conditions [16]. The occurrence of similar meteorological conditions between different locations is phase shifted [16] due to prevailing air flow with a finite speed along slopes of the Transantarctic Mountains.

There is one more important historical piece of additional evidence which confirms my conclusion of temperature data replacement by Captain Scott. The First Relief Party [18], Apsley Cherry-Garrard and Demetri Gerof, was dispatched by Dr. Edward L. Atkinson from Cape Evans on February 26 and returned to the base camp on March 16, 1912. The party arrived at One Ton Depôt in the late afternoon of March 3 by means of dog sledging. They camped there until the morning of March 10. Temperature data were recorded by Cherry-Garrard (usually three measurements per day) using one of the expedition dry-bulb thermometers and are compiled in Simpson's book. Bearing in mind that Captain Scott's party was moving at a speed of ~16 km per day toward One Ton Depôt one can estimate that between March 3 and 10 both parties were separated by a rough maximum of 200 km to a minimum 90 km, respectively. On Figure 3 I have already illustrated that there is no considerable difference between minimum temperatures recorded at the Elaine and Schwerdtfeger stations. The difference, averaged over a period of fourteen years, between near surface minimum air temperature calculated from automated weather stations for the period March 3 and 10, is very small and about -1°F. Since Captain Scott's party was marching in essentially a straight line between these modern weather stations, the differences between the lowest temperatures recorded by Captain Scott and Cherry-Garrard should be negligible; about -1ºF. Contrary to this estimation the average difference in temperatures calculated for seven consecutive days (March 3-10) over a 14 year period is much higher and about -10ºF.

---
[2]In the recently edited journals of Captain Scott [9], the editor, Max Jones while commenting on this particular entry makes a two-fold mistake. The first being that he attributes the four-day blizzard to "a tongue of warm, wet air from the ocean pushing unusually far across the barrier" pp. 499. The second mistake is that he attributes this observation to Solomon [19] who inferred this observation from Schwerdtfeger's book *Weather and Climate of the Antarctic*, reference 73, Chapter 7 of [19]. It was Simpson who pointed out that on December 5, 1911 "the Barrier was affected by a deep depression which appears to have caused a great inflow of moist warm air from the Ross Sea into the south of the Barrier." ([18], vol. I, pp. 28). However, close examination of pressure distributions and temperatures at this time[18] does not confirm Simpson's explanation. Additionally, Fig. 3 of the present paper raises additional questions. Research into this question is needed. On this particular occasion the blizzard conditions were caused by an interaction of Barrier winds along the Transantarctic Mountains, a cyclone over the Ross Ice Shelf or the Ross Sea and katabatic (*Föhn*) wind. For more information see M. W. Seefeldt, *An analysis of the Ross Ice Shelf low-level wind field using surface observations and modeling studies*, Ph. D. thesis, Department of Atmospheric and Oceanic Sciences, University of Colorado, 2007 and references therein.

It was Leonard Huxley who acted as editor and arranged the first edition of Captain Scott's journals [14]. During editorial work Huxley consulted members of the expedition but it was chiefly Cherry-Garrard who acted as a main consultant of added notes [9, pp.313]. The role of Kathleen Scott, Clements Markham and James M. Barrie in getting the journals published was indispensable. The first edition was published in London by Smith, Elder, and in the US by Dodd, Mead and Co. in 1913. To make his case Jones in 2003 observed that "it is now commonly believed that an establishment conspiracy covered up Scott's failings, creating a hero by the careful editing of his sledging journals…" [9, pp.121] Two years later, Jones presented an annotated new edition of Captain Scott's journals together with his comparative analysis of the original journals with the first book's edition. His "systematic comparison of the published journal with Scott's original, exposes relative conventions of the popular biography, not an establishment conspiracy."[9, pp.123]

Table 1. Original temperature recordings form Captain Scott's journal, the changed values published in the first edition of the journal [14] and Dr. Simpson temperature data [18] taken from Lt. Bowers log.

| Date | Scott's entry [ºF] | Changed to [ºF] | Simpson Data [ºF] |
|---|---|---|---|
| 3 November/1911 | +22 | -22 | 5.3, 20.3, 2,7 |
| 7 November | +10 | -10 | 1.7, 9, 10.6, 7.7 |
| 9 November | +12 | -12 | 11.5, 13.1, 1.7, 0.5 |
| 12 November | +10 | -10 | 8.6, 12.1, 10.3, 7.4 |
| 13 November | +10 | -10 | 5, 15.5, 5.6, 2.2 |
| 25 November | +2 | -2 | -9, 0.6 |
| 17 December | +12 | -12 | 12.0, 11.3, 13.5, |
| 18 December | +11 | -11 | 13.3, 12.6, 10.5 |
| 19 December | +11, +5 | -11, -5 | 9.2, 14, 10.8 |
| 28 December | -6 | no entry | -11.1, -7, -6 |
| 11 February/1912 | +6.5, +3.5 | -6.5, -3.5 | 0.7, 6.0, 3.0 |
| 14 February | 0, +1 | 0, -1 | 7.4, 6.6, 0.8 |
| 15 February | +4, +10 | -4, -10 | 1.9, 10.4, 3.5 |
| 16 February | +6.1, +7 | -6.1, -7 | 5.9, 5.6, 6.5 |
| 25 February | +9.5, -11 | -9.5, -11 | -21.7, -10.2, -20.4 |

One of these changes was concerned with discrepancies between temperature entries in Captain Scott's journal and its first printed edition [14]. Jones suggested that "The arbitrary nature of the changes [temperatures] suggests they may have been genuine typographical errors."[9, pp. 457] However, Jones observation is questionable.

During the *Terra Nova* expedition Captain Scott did not record temperatures or another meteorological parameters in any systematic way, as it was not his duty. According to Jones' count [9, pp.457] in the period 18 October 1910 - 29 March, 1912 there were some 175 temperature recordings by Captain Scott. Out of these 175 temperature recordings present in Captain Scott's journal, the sign of 33 temperature readings as printed in the first edition was changed from + ºF to - ºF, excluding the two entries from the 14 and 25 February, Table 1. If these 33 changes of recordings were "genuine typographical errors" as suggest Jones, then one should observe a random distribution of 33 changes in value and time. None of this however is observed.

If the value changes were random, one would expect their average value to be ±0ºF. However on the average (for *each* temperature entry) the temperature was *lowered* by -15ºF. This is contrary to Jones' ruling that, "The reduction of temperatures exaggerated the difficulties faced by the explorers, although the rationale behind the changes is unclear, as most were so small as to be of little significance."[9, pp. 123]

Examining randomness of change frequency leads me to the same conclusion. Out of these 33 changes, only 4 changes were made from {25 Apr. – 31 May} and 29 changes {7 Nov. – 21 Feb.}, respectively. That is 88% of all temperature changes were made during the South Pole Main Party journey. Cleary not a random (in time) distribution.

I believe that the above simple analysis clearly shows that the temperature changes made in the first edition of Captain

Scott's journals were not random (in time and value) and were not genuine typographical errors. They were deliberate changes introduced to dramatize weather conditions encountered by the South Pole Party. It is interesting to notice that after February 25, 1912 no change of temperature entries in Captain Scott's journal was made.

Dr Simpson in his meteorological treatise [18], which I have extensively used in this work, did honestly report with unbiased scientific accuracy [21] the true temperature readings reported by Captain Scott's Main Polar Party in his journals. His were honest readings, not those groundlessly changed under Huxley's supervision. It was not the first time when various false authorities at the Royal Society and Meteorological Office interfered with Captain Scott's field data. A similar situation took place after the *Discovery* expedition of 1901-1904 and not long before the *Terra Nova* expedition. Lt. Charles W. R. Royds was erroneously and groundlessly charged by the scientific establishment with not reporting the true direction of the winds on his South-East Barrier journey [17].

The Main Polar Party used high quality thermometers calibrated at the Kew Observatory, London. Sling and dry bulb thermometers were used with precision. They measured to about a ±0.5ºF certainty [18]. The dry-bulb thermometer was broken by Lt. Bowers [18] on March 10, 1912. From this day on only Captain Scott's spirit thermometer data are available. Simpson [18] informs us that "Nearly all the sledging thermometers (spirit) were provided with minimum indices". However, temperatures measured and reported by Captain Scott after March 10, 1912 *are not* daily minimum near surface temperatures as incorrectly suggested by Solomon[19]. These measurements were taken by Captain Scott on or about the midday and were *actual* measured values at that time. While discussing the mean temperatures, Simpson notices that "The daily range of the temperature on the Barrier is [so] great". Further by using the simultaneous temperature measurements of Bowers and Meares, Simpson observes that the daily (24h) temperature variations are important in calculations of the mean temperatures. On Figure 5 [18, Vol. I, pp.20] Simpson shows that the difference between midday and midnight temperatures on the Ross Ice Shelf (the Barrier) in November 1911 was about -20ºF. In contemporary usage one would say that Simpson was de-trending data to show actual temperature fluctuations.

Returning to analysis and in particular to Fig. 5 one should observe that the temperatures reported by Captain Scott after March 10, 1912 are not actual minimum near surface recordings but the midday measurements[14]: March 13, -37ºF (morning), March 14, -43ºF (midday), March 17, -40ºF (midday), March 18, -35ºF (midday?) and March 19, -40ºF (day time?)[14]. Although I have depicted the just mentioned Captain Scott temperature recordings on Fig. 6, one should readily notice that on the contrary to all shown temperatures, these records are not *minimum* near surface temperatures. This simple fact was overlooked by Solomon who on figures [19, Fig. 3, and Fig. 61] showing daily minimum temperatures, presents a mixture of lowest recorded temperatures and on the contrary to the popular belief of avoiding bias Solomon erroneously observed that, "It is probable that the true 1912 minima lie between the ventilated and under-sledge data" [19, pp. 13014]. True daily minimum temperatures as described by Dr. Simpson [18, vol. III, Table 72, pp. 618] were measured by a spirit thermometer, and, "The party had been taking observations three times a day and had used a minimum thermometer under a sledge to obtain a night temperature."[18, vol. I, pp. 20]

It was Dr. Simpson [18, Vol. I, pp. 20] who analyzed daily variation of temperature at the Barrier and who pointed out that it is vital to account for a 24 h temperature dependence and that the minimum temperatures usually occurred around midnight. Although this relationship is not universal, by taking modern temperature records at the Schwerdtfeger station and by using polynomial fitting of these data, I have estimated that the average difference between midday and midnight temperatures in March is about -12.5ºF. If I accept this at face value, then all temperature records of Captain Scott after March 10 with the exception of March 13, and possibly March 19, the correction of about -12.5ºF must be introduced. Thus, according to Captain Scott the average minimum near surface air temperature in the period of March 10 - March 19, 1912 was more than dramatic -50ºF.

## 5. Concluding Remarks

Captain Scott and his companion's fight against nature and personal weaknesses is well known and has been told with great eloquence many times. My present contribution limits analysis to the investigation of minimum near surface temperatures. The writings on Captain Scott's expeditions are dotted with many unsubstantiated statements and conjectures. I will not comment on all of them, but instead only remark on the information provided by Solomon and Stearns [19], which is directly related to the present work. We all used the same meteorological temperature data, though I used data with an additional ten year record. Solomon and Stearns results [19] feature essentially two points. First, that the temperature data reported by Captain Scott and his party from February 25 until March 19, 1912, are on an average of 10 to 20 degrees below respective modern data (1985-1999). The second point is that only one year (1988 out of the 1985-1999 period) displayed minimum temperatures close to the 1912 reported temperature.

I have to some extent confirmed the first observation by Solomon and Stearns. More precisely the averaged minimum temperature calculated from Captain Scott's historical data is -39.3ºF while for modern data it is -26.0ºF, for February 27 till March 19. The difference is about 13.3ºF. However, I dispute the second observation. It is an observation which has led Solomon and Stearns and many readers; especially the readers of the book *The Coldest March*, to the conclusion that the unusually low temperatures reported by Captain Scott actually have been likely to occur because comparable low averaged minimum near surface temperatures were also observed at least once in 1988 [19].

Fig. 3 combined with Captain Scott's data from Fig. 6 does not confirm this observation. Recasting present data and analysis in terms of Solomon and Stearns approach on Fig. 7 I have depicted averaged minimum near surface temperatures recorded at the weather stations. Analyzing this figure one can readily notice that there are two years (1988 and 1995) which were exceptionally cold in comparison. It appears that at all weather station locations, the average minimum near surface temperature was the lowest in 1988. This temperature which reached -37.5 °F and -36.2°F in 1995[3] was recorded at the location which was the nearest in proximity to Captain Scott's last camp. These averaged temperatures are "close" to the, -39.3°F averaged minimum near surface temperature reported in 1912 by Captain Scott. It appears that subsequent Figs. 4 and 62 of the Solomon and Stearns contributions depict an erroneous value (circa -37°F instead of -39.3°F[18]) of historical (Captain Scott's) minimum temperature data ("with under-sledge data"). This miscalculation, led these authors to a faulty reasoning that their results confirmed and validated the possibility of rare meteorological conditions encountered by Captain Scott and his party. In an indirect way this miscalculation confirmed Captain Scott's declaration that "…no one in the world would have expected the temperatures and surfaces which we encountered at this time of the year" and "the Barrier could be traversed many times without again encountering such low temperatures so early in the year." One should also notice that Solomon and Stearns also wrongly calculated the lowest temperature. They did not take into account the minimum temperature recorded by Scott's party, see Figs. 4 and 62 of Solomon, respectively. This temperature should be -32.8°F, instead of the -34.2°F set by Solomon and Stearns.

However, the main argument of Solomon and Stearns was, that if the likelihood of low temperatures can be confirmed by modern automated weather station measurements then it is certain and/or very likely that the equally low temperatures reported by Captain Scott were indeed observed in the proximity of One Ton Depôt in 1912. If I extend the observation period from ten years to twenty five years, I can then confirm that at least twice on modern record the minimum near surface temperatures were close to that reported by Captain Scott. However, my point which I have shown in this paper is that such an observation does not permit one to claim that Captain Scott and his party really encountered unusually low temperatures in 1912.

I have shown beyond a doubt that there are very high correlations between minimum near surface temperatures at the Ross Ice Shelf. Figure 7 additionally shows and confirms a mirrored similarity of temperatures recorded at the weather stations. All my observations and conclusions were drawn because I have observed a mirrored similarity of temperatures between the McMurdo and Schwerdtfeger stations (One Ton Depôt). What I have established is that even if the temperature at Schwerdtfeger (One Ton Depôt) was unusually low, this unusual decrease was followed by an equally unusual drop in the temperature at McMurdo (Cape Evans). For the 1986-2009 studied period, the lowest recorded minimum near surface temperature was observed in 1988 at the Schwerdtfeger station (-37.5°F). Its gradient was tracked in 1988 by the lowest temperature recorded at McMurdo (-14.5°F, see also Fig.7), and *vice versa*. One should also note that the second lowest minimum temperature occurred in 1995 and is recorded for three stations in my study.

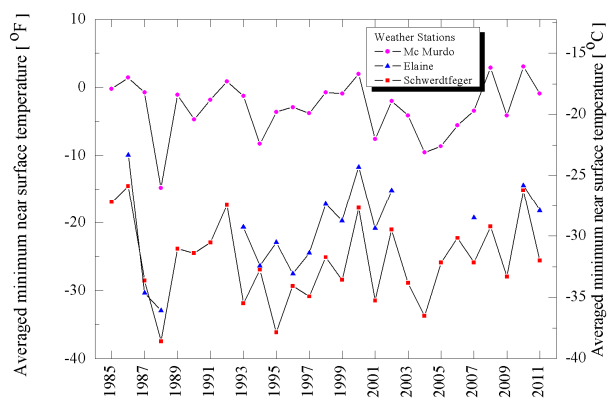

Figure 7. Averaged from over the period of 1985-2009 McMurdo, Schwerdtfeger and 1993- 2009 Elaine from February 27 to March 19 minimum near surface temperatures for three different weather stations on the Ross Ice Shelf.

In the period of February 27 - March 19 the difference between averaged minimum near surface temperatures at McMurdo and Schwerdtfeger stations (1985-2009) is about -22.7°F. The respective difference of averaged minimum near surface air temperatures reported by Captain Scott (in the proximity of One Ton Depôt) and Simpson (at Cape Evans) is -35.0°F. The difference of about 12.5°F between historical and modern *averaged* near surface minimum temperatures at McMurdo and One Ton Depôt is indeed a very big and extraordinary difference.

There is a widely spread and unsubstantiated belief among some polar enthusiasts, historians and hagiographers that the year 1912 was an exceptionally cold one. Due to that it was presumed that extremely low temperatures were indeed encountered by Captain Scott and his party. I have already shown that the temperatures in 1912 were normal at the Ross Ice Shelf and Cape Evans. Restricting analysis to the February 27 - March 19 period one can obtain the following average *minimum* near surface temperature at the base camps of Ross Island for different expeditions (*Discovery*, *Nimrod* [10] and *Terra Nova* [18]): +6.2°F (1902), -6.8°F (1903), +8.5°F (1908), -1.3°F (1911), and -4.3°F (1912). The conclusions are rather palpable: the year 1912 was not a particularly cold one and the year 1903 during the Discovery expedition was the coldest one. It appears that this forgotten truth was well known for years [22]. In fact, if one compares these historical averaged minimum temperatures with the modern average of -3.3ºF (at McMurdo) for the same period of time then one could say that the explorers of the Heroic Age at the Ross Ice Shelf enjoyed quite comparably high temperatures, on the average of +0.6ºF. My analysis of yearly changes of minimum near surface temperatures for modern data reconfirms that temperatures for the year 1912 were usual within the normal fluctuation envelope.

---
[3] Only three minimum temperature data are available at Schwerdtfeger station in 1995 and during the period 27 February and March 19.

Some 30 years ago, Tversky and Kahneman in an influential paper [23] discussed the heuristics and biases that form the foundation of human probability reasoning and judgment. They pointed out that people have an erroneous intuition about the laws of chance [8]. Their thesis although not precise was simple: "People's intuitions about random sampling appear to satisfy the law of small numbers, which asserts that the law of large numbers applies to small numbers as well" [23]. How then would the belief of a "hypothetical scientist who lives by the law of small numbers" [3] affect his statistical interference? It is know from the law of large numbers and central limit theorem, that a large sample is more likely than a small sample to approximate the true distribution of random events. This would be true, for example, of temperature fluctuations for a given time of the year. However, people display early (premature) confidence when streaks occur in random process. The events of random streaks - a low probability event - are likely to happen over and over in time. However, statistical inference based on the occurrence of these small number of events is faulty, as it does not represent the distribution of the underlying population of events. More importantly, people tend to think of streaks as anomalies and suspect that a non-random process is interfering. The most famous and celebrated example of fallacy based on the law of small numbers is the Gambler's fallacy. It is a widely know logical fallacy. This fallacy consists of a bias in which individuals make an inference about future random streaks based on the outcome of previous streaks: the history of streaks will affect future streaks. Usually the Gambler's fallacy is illustrated as repeatedly flipping a (fair) coin and guessing the outcome (streak). Recently, detailed studies of retrospective Gambler's fallacy were conducted [11]. It was shown that "an event that seems rare appears [to the observer] to come from a longer sequence than an event that seems more common" [11]. That for a single rare event, the appearance of rarity is more important than actual rarity. The law of small numbers by its hidden appeal to the law of large numbers, suggests that individuals will not believe that streaks (rare events) should occur in small samples of stochastic processes. Therefore, the individual presented with a rare event has to consider the alternative: either abandon the notion that the process is stochastic, or abandon the notion that the sample is small. Oppenheimer and Monin [11] showed that "people will believe that the series of events have been occurring for a longer time after witnessing a seemingly unlikely event, than if no such event is observed."

I have shown above that Solomon's argument has two sides. First, temperature data from Schwerdtfeger automated weather station to calculate the average minimum temperature at this station in the period from February 25 until March 19, for each year of the record, can be used. Then it can be observed that: "Only 1 year [1988] in the modern record seems to rival the severity of the temperatures measured in 1912" [19]. Let us call the rare events observed in 1988 and recorded in 1912 year as: streak-1988 and streak-1912, respectively. Solomon's argument was that because the streak-1988 was observed then the streak-1912 was also observed by Captain Scott's party. Now, it is self evident that Solomon as well as those readers who were persuaded on this matter by her book and article fell into the trap of the retrospective Gambler's fallacy. There is no logically sound argument that just because streak-1988 occurred, streak-1912 also occurred. Evidently, for Solomon as well as for a good number of readers, their intuitions about random sampling appear to satisfy the law of small numbers, which asserts that the law of large numbers applies to small numbers as well. However, this is logical fallacy.

## 6. Summary

The weather patterns in the Antarctic are neither completely regular nor completely irregular. I have not arrived at conclusions by analysis of logical structures of particular weather descriptions from original journals. In this contribution I have arrived at the conclusions presented because I have rigorously and scientifically analyzed modern and historical weather data. All modern meteorological data, as well as historical data of Cherry-Garrard, high correlations between temperatures at different locations at the Ross Ice Shelf (mirrored similarity), and precision retrodiction of modern and historical data all point out the oddity of Captain Scott's temperature recordings from February 27 - March 19, 1912. On the basis of the mentioned evidence I concluded that the actual minimum near surface temperature data was altered by Lt. Bowers and Captain Scott to inflate and dramatize the weather conditions. Therefore our understanding about the decisive causes of the Main Polar Party deaths must be entirely reassessed. I conclude that their deaths (Scott, Wilson and Bowers) were a matter of choice rather than chance. The choice was made long before there was an actual end of food, fuel and long before the end of physical strength needed to reach imaginary and delusive salvation at One Ton Depôt. I am not authorized to comment on possible events beyond the scientific disciplines presented in this paper.

Finally, while arguing deliberate distortion of temperature data by Captain Scott's Main Polar Party I actually speak here about the truth in a metalogical sense of the *de dicto* sentence about the truth of probability, and not about truth in *de re* sense.

# Acknowledgements

I wish to thank Dr. Leszek Kułak from the Department of Theoretical Physics and Quantum Informatics at the Technical University of Gdańsk, Poland for assistance with neural network simulations. I also wish to thank him for the never ending discussions about entangled issues of the first principles of quantum mechanics. I am also grateful to former and present students: Michał Wołowicz (M.Sc.) for his handling of weather data and Natalia Kułak-Malińska, Sever Niedźwiecki for drawing Fig. 1 and 2, respectively. I thank those at the Antarctic Meteorological Research Center, University of Wisconsin, USA (Linda M. Keller) and the British Antarctic Survey, UK (Steven R. Colwell) for their archive of Antarctic automated weather station data.